\documentstyle[osa,manuscript,psfig]{revtex}

\begin{document}

\titlepage
\title{
Colorless and colored gluon-clusters in nucleon?}
\author {C. Boros, Liang Zuo-tang and Meng Ta-chung}
\address {Institut f\"ur Theoretische Physik,
Freie Universit\"at Berlin \\
Arnimallee 14, 14195 Berlin, Germany}
 
\maketitle

\begin{abstract}                

It is suggested that virtual gluon-clusters exist in
nucleon, and that  such colorless and colored 
objects  manifest themselves in  the small $x_B$
region of inelastic lepton-nucleon scattering processes. 
The relationship between the space-time properties 
of such clusters and the striking features observed 
in these scattering processes is discussed.  
A phase-space model is used to show how quantitative results 
can be obtained in such an approach. 
The results of this model-calculation are in reasonable 
agreement with the existing data. 
Further experiments are suggested.

\end{abstract}

\newpage
 
High precision data for deep-inelastic electron-proton  
scattering  and photon-proton scattering in the  small $x_B$
(down to $1.8\times 10^{-4}$)
region are now available$^{/1,2/}$ from HERA. 
These data not only 
show that in this region, $F_2(x_B,Q^2)$ 
increases with decreasing $x_B$ at fixed $Q^2$, 
and increases like $\ell n Q^2$ at fixed $x_B$, 
but also show that a distinct class of events ---
the large rapidity gap events --- exist.
These striking features of the data, 
have already generated much 
interest$^{/3/}$. 
Are they related to one another ? 
Can these observations shed light in understanding the structure of nucleon, 
and/or the reaction mechanism(s) of such collision processes ?  

It is known$^{/1,2,3/}$ that the data 
for large rapidity gap events
can be described$^{/4/}$ 
in terms of the special Regge-pole, 
Pomeron, and/or other Pomeron-like
objects, all of which are supposed 
to have parton-structures.
It is also known$^{/5/}$ 
since the 1970's that Pomeron  
may be considered as a ``gluon-system''. 
We are thus led to the following questions: Are these results suggesting 
that independent  gluon-systems exist in proton ? 
What properties, in particular  space-time properties,  
should such  systems --- Pomeron and/or Pomeron-like objects ---  
have ? According to  Regge Pole Models,  the  Regge-pole Pomeron, is 
associated with vacuum 
quantum-numbers; hence it is expected that these objects should  be 
colorless and carry no isospin etc. But gluons are members of
color-octet; hence a system of gluons should in general carry color. 
Can corresponding colored systems also exist in 
proton if colorless ones do ? If no, why not ? If yes, 
how do they manifest themselves experimentally? 

In order to answer these questions, it seems useful to recall 
the following: 
The proton is a spatially extended object, 
and by performing deep-inelastic electron-proton scattering
in the small $x_B$ and large $Q^2$ region,
we wish to probe the structure of the proton  by a beam of
incoming 
virtual photons $\gamma^*(x_B,Q^2)$ of small transverse spatial
dimension
($\sim 1/Q^2$) and of relatively short interaction-time 
(which is equal to 
$1/\nu \equiv 2Mx_B/Q^2$ when viewed from its rest frame, 
and it is  
$4P/(2M\nu-Q^2)$ when viewed 
in the electron-proton cms system where $P$ is the
momentum the proton).
In other words, in carrying out such experiments, we are asking
ourselves: 
What kind of information about the proton structure can we obtain by
using virtual photons
$\gamma^*(Q^2,x_B)$, with  (transverse) 
resolution power $1/Q^2$, which interact
with the constituents of the proton for a short  time-interval 
($1/\nu\equiv 2Mx_B/Q^2$ or $4P/(2M\nu-Q^2)$ in the reference 
frames mentioned above),  
 provided that $\gamma^*(Q^2,x_B)$ is absorbed 
by the constituents of the proton, before it 
turns itself into lepton- or quark-pairs due to vacuum-fluctuation. 
Can such virtual photons tell us in particular whether (if yes how) the
constituents of the proton interact with one another? 

First of all, let us recall that, 
not only the spatial resolutions power, but also 
 the interaction-time is important in studying this problem: 
In the limiting case, in which the interaction-time between 
the virtual photon and  
a (or a group of) constituent(s) is much longer than the average time
for color-interaction to propagate between any two constituents of the
proton, the information we get from ``our messenger'', the virtual photon,
will be: 
``Every constituent of the
proton is
directly or indirectly interacting with every other constituent of the same
proton.''.
Hence we are forced to say that, in this case, all constituents 
are interacting with one another. 
But, we are also forced to accept that 
{\it not all} constituents inside the proton are
interacting with one another {\it in a given time interval} ---
especially when this time-interval is much shorter than the average
time for color-interaction to propagate between any two constituents.
This is because in the other limiting case, in which the
interaction-time between the virtual photon and the struck 
constituent (of the proton) is extremely short 
--- so short that it does not have time to communicate with any other 
constituent of the proton,  
 every struck constituent (parton) can be considered as a free 
particle during the interaction-time. This is the 
well-known result of Parton Model and/or impulse-approximation, which
explain(s) the existence of approximate Bjorken-scaling. 

In this connection, it is also useful to recall the following: 
Due to vacuum polarization in QED, 
the virtual photon may dissociate into a
fermion-antifermion pair 
(lepton-pair or quark-pair) which exist for a certain time-interval; 
and the question whether the virtual photon reaches the struck
constituent of the proton in form of a bare photon or in form of a 
fermion-pair, depends on the life-time of the virtual fermion-pair state. 
This life-time can be calculated by standard methods with the help 
of the uncertainty principle. 
Viewed from proton`s rest frame, it is of order $q_{||}/m_{\perp}^2$ 
where $q_{||}$ is the longitudinal momentum of the photon, and 
$m_{\perp}\equiv (m^2+p_{\perp}^2)^{1/2}$ is the so-called transverse
mass of the fermion 
(with $m$ stands for the mass
of the lepton or that of the quark, and $p_{\perp}$ its transverse
momentum).  
It means, viewed in this reference frame, the life-time of the virtual 
fermion-antifermion pair state created by an energetic photon is long
--- much longer than the interaction-time between 
a photon (with the same energy) and 
(the constituents of) the proton. Does this mean that the ``prober'' 
is very likely {\it not} a bare photon?  
The difficulty is bypassed in Parton Model by describing 
the electron-proton scattering in a ``fast moving frame'' 
in which the proton ( and thus all its 
constituents) are moving with light-velocity toward 
$\gamma^*$. We note that, this method of bypassing the above-mentioned 
difficulty has also been used in all 
the ``QCD-corrected parton models'' (See e.g.  Refs 1-3 and 
the references given there). In fact, all the parton 
(valence quark, sea quark and gluon) distributions extracted 
from the data are based on theoretical 
interpretations and/or analyses made in reference-frames in which 
the constituents are moving with (almost) 
light-velocity. 

Viewed from such a fast moving reference frame, a gluon $g$  
of the  proton  
may dissociate into a quark-antiquark pair, and the 
life-time of such a virtual quark-antiquark $q\bar q$  
 state is $\tau_g \sim q_{g||}/m_{\perp}^2$ 
where $q_{g||}$ is the longitudinal momentum of the gluon $g$ and 
$m_{\perp}$ is the transverse mass of the quark $q$  (antiquark
$\bar q$). Hence, if 
$\tau_g$ is longer than the above-mentioned photon ($\gamma^*$) 
interaction-time $\tau_{int}$, 
this virtual $q\bar q$ state
can be detected by the  $\gamma^*$ with the given interaction-time 
$\tau_{int}$.  
Now, $\gamma^*$ can only interact with charged
constituents, a gluon can dissociate into a virtual quark-antiquark 
pairs, and the life-time of such a virtual $q\bar q$-pair is directly 
proportional to the longitudinal momentum of the gluon. 
These facts might lead us to 
conclude  that, for a  photon $\gamma^*$ with given 
interaction-time $\tau_{int}$,
only gluons with sufficiently large longitudinal 
momenta could be (indirectly) detected by this $\gamma^*(\tau_{int})$. 
Can gluons with lower longitudinal momenta also have a chance 
to be (indirectly)  detected by such a $\gamma^*(\tau_{int})$ ? 

The answer is ``Yes!''. This is because 
according to QCD, quark-antiquark pairs  
can also be produced 
by gluon-gluon collisions,  
and the produced quark-antiquark pairs may again turn into gluons. 
One of the simplest examples is the process illustrated by the so-called 
``box-graph'' (See Fig.1) in which two 
gluons turn into a quark-antiquark pair and than turn into two gluons. 
Now, imagine that, in a certain time-interval, 
two gluons interact with each other 
(via quark-exchange) and create a quark-antiquark pair which
after 
a short moment turn into two gluons again; and 
during this short moment, a virtual photon $\gamma^*$ enters 
and interact with the  quark or the antiquark for a time-interval
$\tau_{int}$. Our messenger $\gamma^*(\tau_{int})$ would report: 
``I see charges distributed in space-time! '', and we know, because of
energy-momentum conservation (and uncertainty principle),  
 the life-time of such a virtual quark-antiquark 
state is proportional to the sum of the longitudinal momenta of these 
two gluons. 
This example explicitly demonstrates how two gluons may come together 
to form a quark-antiquark pair, which can be detected by virtual 
photons with a fixed interaction-time.  
If these two gluons, which form the quark-antiquark 
pair, for a given time-interval do not interact with other constituents
of the proton (See the discussion at the 
beginning of the present paper!), we call this system ($c^*$) of gluons 
``a gluon-cluster''.  
We call the time-interval in which they do not interact with other
constituents of the proton 
``the life-time of this gluon cluster'', and denote this quantity  by $\tau_c$. 
Since three or more gluons can also form quark-antiquark pair(s), as 
illustrated in Fig.1, while the life-time of such quark-antiquark
pair(s) state is directly proportional to the sum of all 
participating gluons.  
(This can be readily seen by straightforward generalization of the graphs
shown in Fig. 1 to higher orders.) It should be mentioned in this 
connection,  that 
the notion of gluon-cluster is {\it not} restricted to a system of two
or more gluons which can form 
 one and just one
quark-antiquark pair.  
A gluon cluster $c^*$ is in general a system of gluons which,
for a given time-interval $\tau_c$, interact with one another, and only
with one another; they form quark-antiquark pairs,  
and the space-time distributions of the temporarily existing 
charged constituents can be detected
by virtual photons in inelastic electron-proton scattering processes. 
The   life-time of the gluon
cluster $\tau_c$ can be calculated by using the 
uncertainty principle.

Kinematic considerations show that such a gluon-cluster, before
it interacts (i.e. its charged constituents    interact) 
with the incoming virtual photon
$\gamma^*(x_B,Q^2)$, is itself {\it virtual} --- 
in the sense that the total four-momentum $q_c$ of the
system of gluons is such that the scalar-product $q_c^2$ is less
than zero. 
This is  why we denote such a cluster by 
$c^*$ (with an asterisk as superscript). 
It should also be mentioned in this connection that, 
in contrast to gluon-clusters,  the 
four-momentum of  a glueball is 
time-like.  
Perhaps, it is possible to produce glueballs 
and/or mesons by knocking-out 
colorless gluon-clusters under 
appropriate experimental conditions. 

Having in mind that the gluons are subjected to confining color
forces, the magnitude of
which increases with increasing distance within a hadron, and that the  
clusters are 
formed in a random manner, it is not difficult to imagine that the typical
spatial extension of a gluon-cluster is comparable with that of a
hadron, and that 
there are in general spatial overlaps between different clusters of the same
proton. But, as we have explained in connection with 
the definition of the clusters,  
during the given interaction-time
$\tau_{int}$,
the cluster struck by $\gamma^*(Q^2,x_B)$ can be considered as 
a free (that is not interacting with other cluster or hadrons) 
object,  provided that $\tau_{int}$ is shorter than
the life-time of the cluster. 
(As we have already mentioned at the beginning of this paper,  
the question ``who is interacting with whom'' 
depends on the comparison between two time-intervals: 
the time-interval the ``prober'' does his measurement and the
time-interval 
the probed objects need to communicate with one another!) 
Furthermore, since gluons are members of a color-octet, 
gluon-clusters are expected to be either in the color-singlet 
(colorless) or in one of the possible color-multiplet (colored)
states. 
Hence, there should be 
two kinds of gluon-clusters, 
which we denote by $c_0^*$ and $c_m^*$ respectively. 
This means, once we accept the widely accepted 
relationship between color and confinement, 
we also have to accept that
the spatial distribution of $c_0^*$ should be 
very much different from that of $c_m^*$. 
In particular, the former may exist beyond 
the ``average border'' for 
the colored constituents of the proton.   
Taken together the fact that
there is no color-connection 
between the colorless clusters $c^*_0$ and the 
rest of the proton,
this  immediately lead us to  the following conclusion: 
The interactions between the virtual 
photon $\gamma ^*$ with $c_0^*$'s are peripheral 
$\gamma^*$-nucleon collisions; and it is this kind of 
collisions which is responsible for the large rapidity gap events. 
The interactions between the virtual photon $\gamma ^*$ 
with the colored clusters $c^*_m$'s do not 
contribute to the large rapidity gap events, but 
they do contribute to the inclusive cross-section 
for deep-inelastic lepton-nucleon collisions, and hence 
to the structure functions of the nucleon.

Before we proceed to discuss in more detail the relationship between 
 the proposed  gluon-clusters and the existing  
  inelastic electron-proton scattering data,  
it seems useful to ask: 
How much {\it do we know} about the {\it dynamics} of cluster-formation and
cluster-decay ?
How much details {\it do we need to know} about the {\it dynamics} 
of such formation-
and decay-processes  in order to understand
the results obtained the experiments performed at HERA ? 
The answer to the first question is: ``Very little.'' It is 
the case,  although we adopt  
QCD  for the description of the
elementary interactions between the constituents (quark and gluons).
As we can see from the examples shown in Fig.1, the
color-interactions 
between the constituents in a cluster are in general very complicated,
and 
we are not sure whether it is useful and/or meaningful to calculate the
lowest order or certain sets of graphs by using 
perturbative QCD. 
To attack such random formation
process using non-perturbative methods 
in which the interaction-time with the ``prober'' is also 
taken into account, 
lattice QCD calculations seems to be an attractive possibility. Discussions
on the possibility of writing a Monte-Carlo program based on such ideas 
--- in particular the idea of also taking the interaction-time between 
the ``prober'' and the ``probed object'' is taken  
into account in such calculations ---  are 
underway; but we are still rather far from our goals.  
Fortunately enough, the answer to the second question is
{\it also} ``Very little.''! In fact,  
in the present paper, we discuss the problem by using 
a {\it statistical approach}; and we  
show that phase space considerations {\it without any dynamical input}  are 
sufficient to give a reasonable description of the striking 
characteristic features of the HERA-data.

To be more precise, let us 
consider such gluon-clusters  and their
interactions with virtual photons $\gamma ^*$ 
(with given $Q^2$ and $x_B$) in  a fast moving frame 
 in which the proton is moving 
practically with the velocity of light. 
We denote the four-momentum of the proton by $P=(E_p,0,0,|\vec P|)$, 
that of the cluster by 
$q_c=(q_c^0,q_{c\perp}\cos\varphi_c, q_{c\perp}\sin\varphi_c,q_{c\|})$ 
and that of the remnant of the proton by $P'$. 
Here, we consider 
only the case in which 
the remnant of the proton remains  
a proton (the generalization is straightforward)
and obtain that, for $E_p^2\gg  Q^2\gg M^2$ 
($M$ is the proton mass), 
\begin{equation}
\tau_c={1 \over \Delta E}\approx {\xi_cE_p\over Q_c^2},
\end{equation}
where $\Delta E=P'^0+|\vec q_c|-E_p=|\vec q_c|-q_c^0$
is the energy difference between the
initial nucleon and the virtual state
(which consists of the cluster
$c_i^*$ and the proton remnant). 
In the above-mentioned reference frame, the variable 
$\xi_c\equiv (q_c\cdot q)/(q\cdot P)$ is simply the 
fractional energy of the cluster. 

Colored clusters can only exist inside the nucleon, 
that is in a restricted spatial region, 
the length of which is of order  $1/M$ 
(the Compton wavelength of the nucleon). 
This means, due to confinement, the $c_m^*$'s are 
expected to stay in a volume $V_m$ of the order of $M^{-3}$
independent of the four-momenta $q_c$ of the clusters. 
In contrast to this, 
the $c_0^*$'s 
can leave the nucleon 
and exist in a spatial region 
which depends on the spatial distance that  a $c_0^*$ 
can propagate.  
In the infinite momentum frame, the volume of this 
spatial region is given by $V_0=\pi r_0^2 l$. 
Here $r_0\propto \tau_c$ is a measure of 
the transverse distance  
which $c_0^*$ can travel during its lifetime $\tau_c$.  
Since everything practically moves with the 
same (light) velocity in the longitudinal direction, 
$l$ depends, if at all, only weakly on $q_c$ and 
is taken as a constant, $l\sim M^{-1}$, 
independent of $q_c$. 
Hence, we obtain, for the four dimensional 
phase-space $\Omega _i$ for $c_i^*$ with 
$i=0$ and $m$, the following:
\begin{equation}
\Omega_i(\xi_c,Q^2_c)\equiv \tau_cV_i 
\propto \left \{
\begin{array}{ll}
\tau _c^3 M^{-1}, &\mbox {for $i=0$}\\ 
\tau _c M^{-3}, &\mbox {for $i=m$}
\end{array} \right .
\end{equation}

Now, we consider the number-density $N_i(q_c)$ of $c_i^*$
($i=m,0$) with fixed $q_c$. 
It is clear that, 
unless there are special dynamical reasons
which forbid such considerations,
the simplest ansatz for this quantity 
is to assume that it is proportional to the corresponding
allowed (4-dimensional) phase-space $\Omega _i$.
That is:
\begin{equation}
N_i(q_c)=\kappa_i M
\delta (q_c^0-\xi_cE_p)
\Omega _i(\xi_c,Q_c^2)
\end{equation}
Here, the $\delta $-function takes care of  to 
energy-momentum conservation which requires   
$q_c^0  \approx \xi_c E_p$ for 
$E_p^2\gg Q^2\gg M^2$; and
$\kappa _i$ with $i=0, m$ are two constants.

The contributions 
to the nucleon structure function 
$F_2(x_B,Q^2)$,  
$F_{2}(x_B,Q^2|c_m^*+c_0^*)
=F_{2}(x_B,Q^2|c_m^*)+F_{2}(x_B,Q^2|c_0^*)$, 
can be calculated. 
They are given by, 
\begin{equation}
F_{2}(x_B,Q^2|c_i^*)=
\int _{D} d^4q_{c}
N_i(q_{c})
F_{2}^{ci}(q,q_c),
\end{equation}
where  $F_2^{c_i}$ 
stands for ``the structure function of the cluster $c^*_i$''. 

What do we mean by ``the structure function of the cluster $c_i^*$
(i=0,m) ? 
Why do we simply add the contribution from 
$c^*_0$ and that from $c^*_m$ ? These questions are not 
difficult to answer. This is because, 
having the proposed picture in mind, 
what we need to know are the following:  
For a given $Q^2$ and
a given $x_B$, how large is the chance for $\gamma^*(Q^2,x_B)$ to
hit a charged member of a gluon-cluster of life-time $\tau_c$ ?
Now, it is clear that probability for  
such a $\gamma^*(Q^2,x_B)$ to hit
a charged constituent of a gluon-cluster $c_i^*$
(i=m,0) is different from zero if such a cluster exists. 
It is also not difficult to imagine that 
we can express this
probability, in analogy to that 
for a hadron under such circumstances, 
 as ``the structure function of the cluster
$c_i^*$'' which is in general a function of the kinematic
variables $Q^2,x_B,Q^2_c\equiv -q_c^2, x_{Bc}\equiv Q^2_c/(2qq_c)$.
Here, $q$ and $q_c$ are the four momenta of $\gamma^*$ and
$c_i^*$ respectively. In particular, since a $\gamma^*(Q^2,x_{B})$
can hit either a charged constituent of a
$c_m^*(Q^2_c,x_{Bc})$ or that of a  $c_0^*(Q^2_c,x_{Bc})$, the
probability for $\gamma^*(Q^2,x_{B})$ to hit a charged constituent
of {\it either} $c_m^*(Q^2_c,x_{Bc})$ {\it or} $c_0^*(Q^2_c,x_{Bc})$ 
should be {\it the sum}
of the corresponding possibilities.
This is why we add the structure functions of $c_m^*$ and $c_0^*$.  
In this connection, it is also interesting to compare the structure 
function
of a virtual gluon-cluster with that of a virtual meson 
--- in particular  a  virtual pion. The latter is similar to the former 
in the sense that it also  describes  
the probability for the virtual photon in an inelastic 
electron-proton scattering to interact with 
a charged constituent of a virtual 
(that is space-like) subunit of the proton. 
But, in contrast to the former, it is  formed by quark-antiquark 
collisions, and thus manifest itself outside the small-$x_B$
region where the quark/antiquark-distributions dominate. 
This is why, we do not consider them in the present paper.

The physical region $D$ 
of the above-mentioned gluon-clusters  is characterized by:
$q^2_{c\perp }\ge 0,\,\, q_c^0+q^0>0$
and $(q+q_c)^2\ge m_\pi^2$.
Hence, the integration limits, and therefore
the integrals, are functions of 
$x_B$ and $Q^2$ --- 
even when $F_2^{c_i}$ are taken as constants  
(see the discussions below). 
For the explicit evaluation of
$F_2(x_B,Q^2|c_i^*)$
given by Eq.(4),
it is convenient to use,
instead of $(q_c^0,\vec{q_c})$,
the following set of
variables $(q_c^0, \xi_c, Q_c^2, \varphi _c) $.
The integration limits
$\xi_{cmin}(x_B,Q^2)\le \xi_c\le \xi_{cmax}(x_B,Q^2)$ and
$Q^2_{cmin}(\xi_c;x_B,Q^2)\le Q^2_c\le Q^2_{cmax}(\xi_c;x_B,Q^2)$
are determined by the above-mentioned
kinematic constraints.
In order to see the qualitative features of the relevant quantities, it
is useful to know that, for 
$Q^2\gg M^2$,  we have:
\begin{equation}
Q^2_{cmax}\approx Q^2(\xi_c/x_B-1), 
\end{equation}
\begin{equation}
Q^2_{cmin}\approx M^2\xi_c^2/(1-\xi_c),
\end{equation}
and $\xi_{cmax}\approx 1$, $\xi_{cmin}\approx x_B$.

The following qualitative features
can be read off from Eqs.(1-6):
Both $N_i(q_c)$'s ($i=0,m$) depend 
on $Q_c^2$ and $\xi_c$, and both
decrease with increasing $Q_c^2$ 
for fixed $\xi_c$.
For a given $\xi_c$, 
the average number 
$\langle N_i\rangle$ of clusters $c_i$ 
which can be detected by the photon $\gamma^*(Q^2,x_B)$  
can be readily obtained by integrating
$N_i(q_c^0,\xi_c,Q^2_c,\varphi_c)$ 
in the physical domain $D$ over $q_c^0$, 
$Q_c^2$ and $\varphi_c$. 
The results are the following: 
$\langle N_0(x_B,Q^2;\xi_c)\rangle $
is proportional to $\xi_c^3/Q_{cmin}^4$, 
and $\langle N_m(x_B,Q^2;\xi_c) \rangle$ is directly
proportional to $\ell n(Q^2_{cmax}/Q^2_{cmin})$.
Thus we have, 
\begin{equation}
\langle N_0(x_B,Q^2;\xi_c) \rangle \sim {1\over \xi_c},
\end{equation}
\begin{equation}
\langle N_m(x_B,Q^2;\xi_c) \rangle \sim 
\ell n{Q^2\over M^2\xi_c^2}({\xi_c\over x_B}-1),  
\end{equation} 
for $x_B<\xi_c\ll 1$. 
It is interesting to note in particular 
that the latter has a term proportional 
to $\ell nQ^2/M^2$, and this dominates  the $Q^2$-dependence
of $F_2(x_B,Q^2|c_m^*)$. 
The former is expected to have significant 
influence in particular on 
$\xi_c$-dependence of
the ``diffractive structure function''
$F_2^{D(3)}(x_{Bc},Q^2,\xi_c)$
which will be discussed below.

Let us now examine the second factor
in the integrand of Eq.(4),
the structure function of the cluster $c_i^*$, 
$F_2^{ci}$, which is a priori not known. 
As a first step, we ask:
What kind of $Q^2$- and $x_{B}$-dependence
do we expect to see for $F_2(x_B,Q^2|c_i^*)$, 
if the dynamical details of
such a process do not play a role at all?
To see this, we took $F_2^{ci}$ as a constant. 
It is interesting to see that
such a simple ansatz is consistent with the
experimental finding
in $e$-$p$-scatterings$^{/1,2,3/}$
and that in $e^+$-$e^-$-processes$^{/9/}$.
 
In terms of the variables
$(q_c^0,\xi_c,Q^2_c,\varphi_c)$,
we obtain from
Eqs.(2) and (4), with $F_2^{ci}=$const,
the following:
\begin{equation}
F_2(x_B,Q^2|c_i^*)=\lambda \kappa _i
\int _{\xi_{cmin}}^{\xi_{cmax}} d\xi_c
\int _{Q^2_{cmin}}^{Q^2_{cmax}} dQ^2_c \ \Omega _i(\xi_c,Q^2_c),
\end{equation}
where $\lambda \approx 4\pi M/ (1+x_BE_p/E_e)$ 
($E_e$ is the electron energy) coming from the 
Jacobi one obtains from the variable-transformation 
from the four-component of $q_c$ to   
$(q_c^0,\xi_c,Q^2_c,\varphi_c)$ 
and the integration over $\varphi _c$. 
The $\Omega _i(\xi_c,Q^2_c)$'s are given by Eq.(3).
This means, under the assumption that gluon-clusters indeed exists in
the small $x_B$ region, 
and that the dynamical details about such cluster can be 
neglected in a statistical approach, their 
contributions to $F_2(x_B,Q^2)$ namely the sum of 
 $F_2(x_B,Q^2|c_i^*)$'s can now be 
calculated and be compared with 
data$^{/1,2,3,6,7/}$.  
Such calculations have been carried out by using the exact expressions
for the 
kinematic boundaries.  
The two unknown parameters 
$\kappa _i$'s ($i=m,0$)
times the over all normalization constant
are determined by adjusting 
them to the normalization
of the data points.
The results are shown in Figs.2a and 2b.
In order to see whether/how they 
contribute to the large rapidity gap events, 
we have also calculated the rapidity distribution 
of the produced hadrons. 
To do this, we wrote a 
Monte-Carlo program to generate events according 
to Eq.(4) and took the following into account: 
While a $\gamma^*$-$c^*_0$ system 
can fragment independently, 
a $\gamma^*$-$c^*_m$ system (because of the color lines) 
has to fragment together with the rest of the proton. 
The fragmentation was performed by using the Lund model$^{/10/}$ as
implemented in JETSET$^{/11/}$. (The details of this Monte-Carlo
calculation will be published elsewhere$^{/12/}$.). 
The result is shown in Fig.~3.

Furthermore, we note that
the measured$^{/1,2,3/}$
``proton's diffractive structure functions''
$F_2^{D(4)}(\beta,Q^2,x_{\cal {P}},t)$ and
$F_2^{D(3)}(\beta,Q^2,x_{\cal {P}})\equiv
\int F_2^{D(4)}(\beta,Q^2,x_{\cal {P}},t) dt$
can also be calculated 
and that the variables used in Refs.1 and 2
are closely related to those we used in this paper.
In fact, we have: $x\equiv x_B$, $\beta \equiv x_{Bc}$,
$x_{ P}\equiv \xi_c$, $t\equiv q^2_c$;
and in the case $F_2^{ci}=$const, we have, 
\begin{equation}  
F_2^{D(4)}(x_{Bc},Q^2,\xi_c,Q_c^2)=
\lambda \kappa _0\Omega _0(\xi_c,Q^2_c),
\end {equation}
\begin{equation}
F_2^{D(3)}(x_{Bc},Q^2,\xi_c)=
\lambda \kappa _0
\int _{Q^2_{cmin}}^{Q^2_{cmax}}dQ_c^2 \ \Omega _0(\xi_c,Q^2_c).
\end{equation}
They can now be calculated 
without any adjustable parameters 
and the comparison of the obtained results with
the experimental findings 
should be considered as further tests of the
proposed picture. 
In fact, Eq.(10) shows that 
$F_2^{D(4)}(x_{Bc},Q^2,\xi_c,Q_c^2)$
should in particular be proportional to $\xi_c^3$ 
for fixed $Q_c^2$ and proportional to $Q_c^{-6}$ 
for fixed $\xi_c$. 
Eq.(11) implies that
$F_2^{D(3)}(x_{Bc},Q^2,\xi_c)$ should
behave approximately in the same way as the 
$\langle N_0(x_B,Q^2;\xi_c)\rangle $ does, 
namely it should be approximately 
proportional to $1/\xi_c$  [see Eq.(7)].
While the latter is consistent
with the existing data$^{/1,2,3/}$ (see Fig.4),
the former can be checked by future experiments.

In conclusion, it is suggested that 
virtual gluon-clusters exists, and they manifest themselves 
in the  small $x_B$ region of inelastic lepton-nucleon
scattering processes. 
It is shown that the space-time properties of the gluon-clusters  
play a very special role in understanding the striking properties observed 
in such scattering processes. 
To be more precise, 
it is shown that the persisting $lnQ^2$-dependence of $F_2(x_B,Q^2)$ and 
the existence of large rapidity events are closely related 
to each other; in fact they directly 
reflect the space-time properties of the colored and the colorless
clusters. 
In the framework of a statistical approach 
discussed in this paper, the 
gluon-clusters are randomly formed, and the dynamical details of
cluster-formation and 
cluster-decay are completely neglected. In order to make quantitative 
comparisons with the data$^{/1-3/}$,  
a phase-space model is used to
carry out the  calculations. It is seen that the  results of this 
model-calculation are in reasonable agreement with the existing 
data. Further studies along this line will show 
whether/how the dynamical aspects of such clusters can significantly
effect the obtained results.

We thank J.~Bartels, X.~Cai, D.H.E.~Gross, L.~Liu, R.~Rittel,
O.~Shapiro, K.-D.~Schotte and W.~Zhu for helpful discussions.
This work is supported in part by
Deutsche Forschungsgemeinschaft (DFG: Me 470/7-1).

\newpage

\begin{figure}
\psfig{file=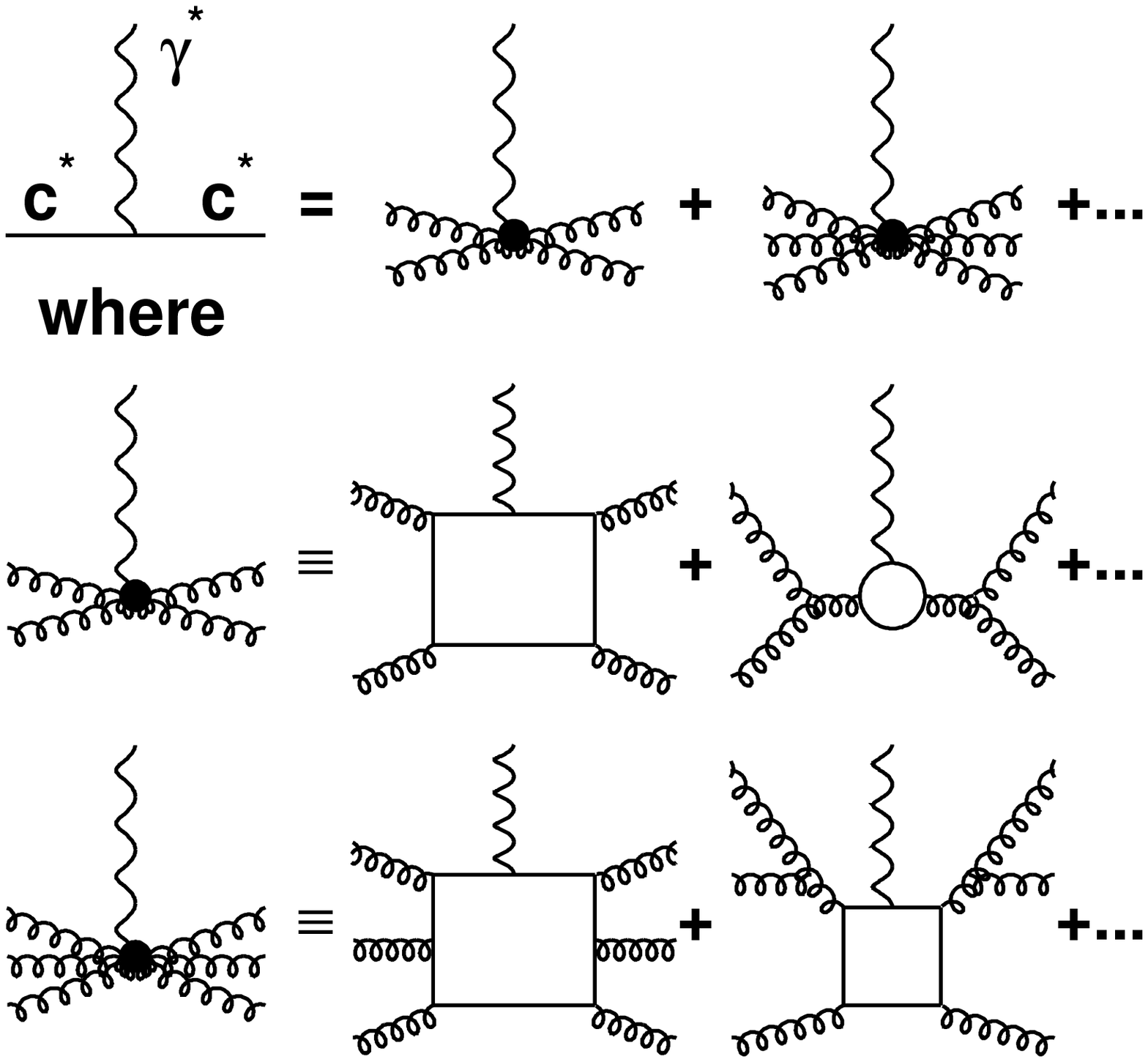,width=13cm,angle=270}
\caption{ An illustrative example: 
a gluon-cluster $c^*$ that absorbs a photon $\gamma^*$.} 
\end{figure} 

\newpage

\begin{figure}
\psfig{file=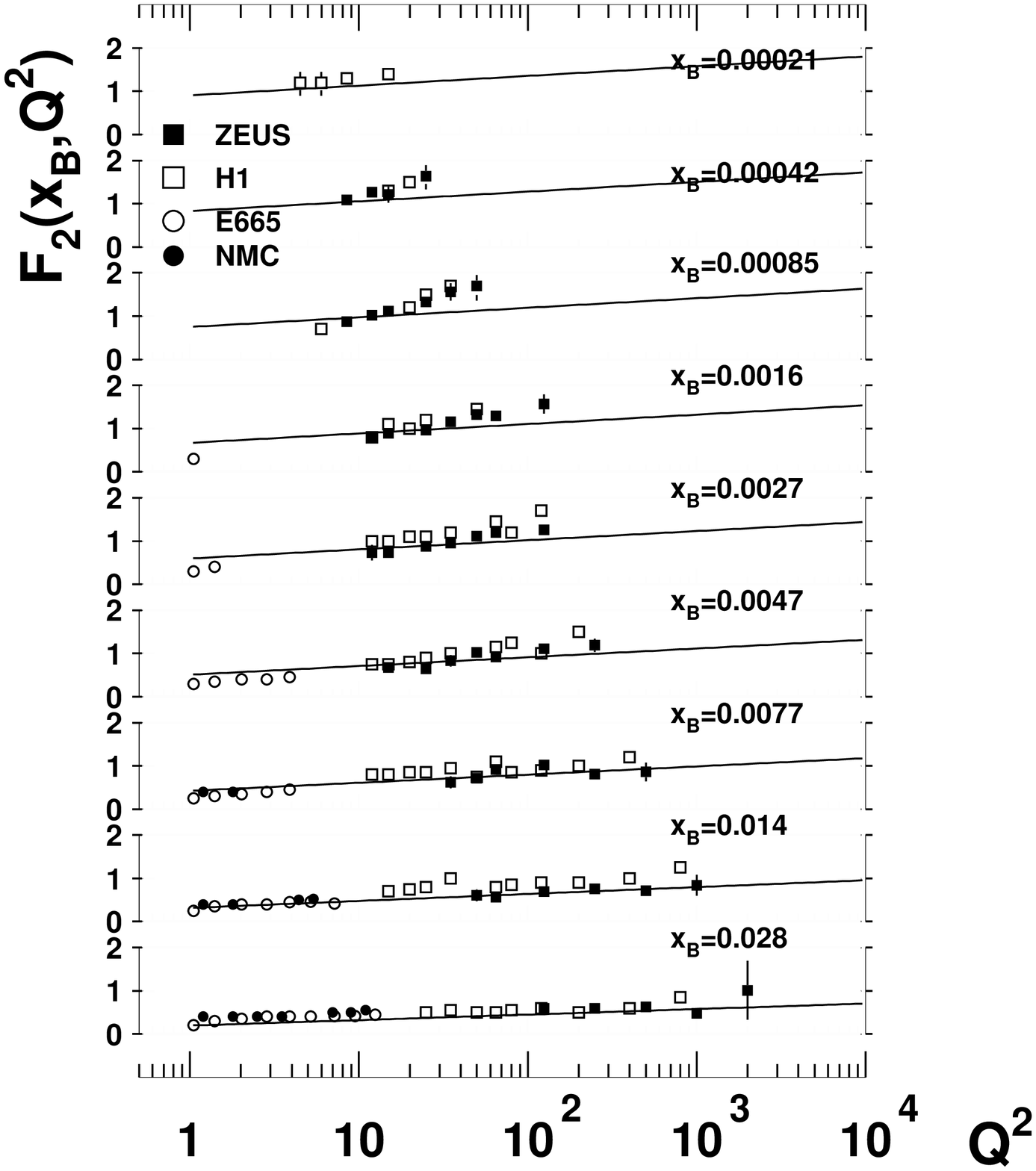,width=8cm}
\end{figure}
\vspace*{-1.5cm}
\begin{figure}
\psfig{file=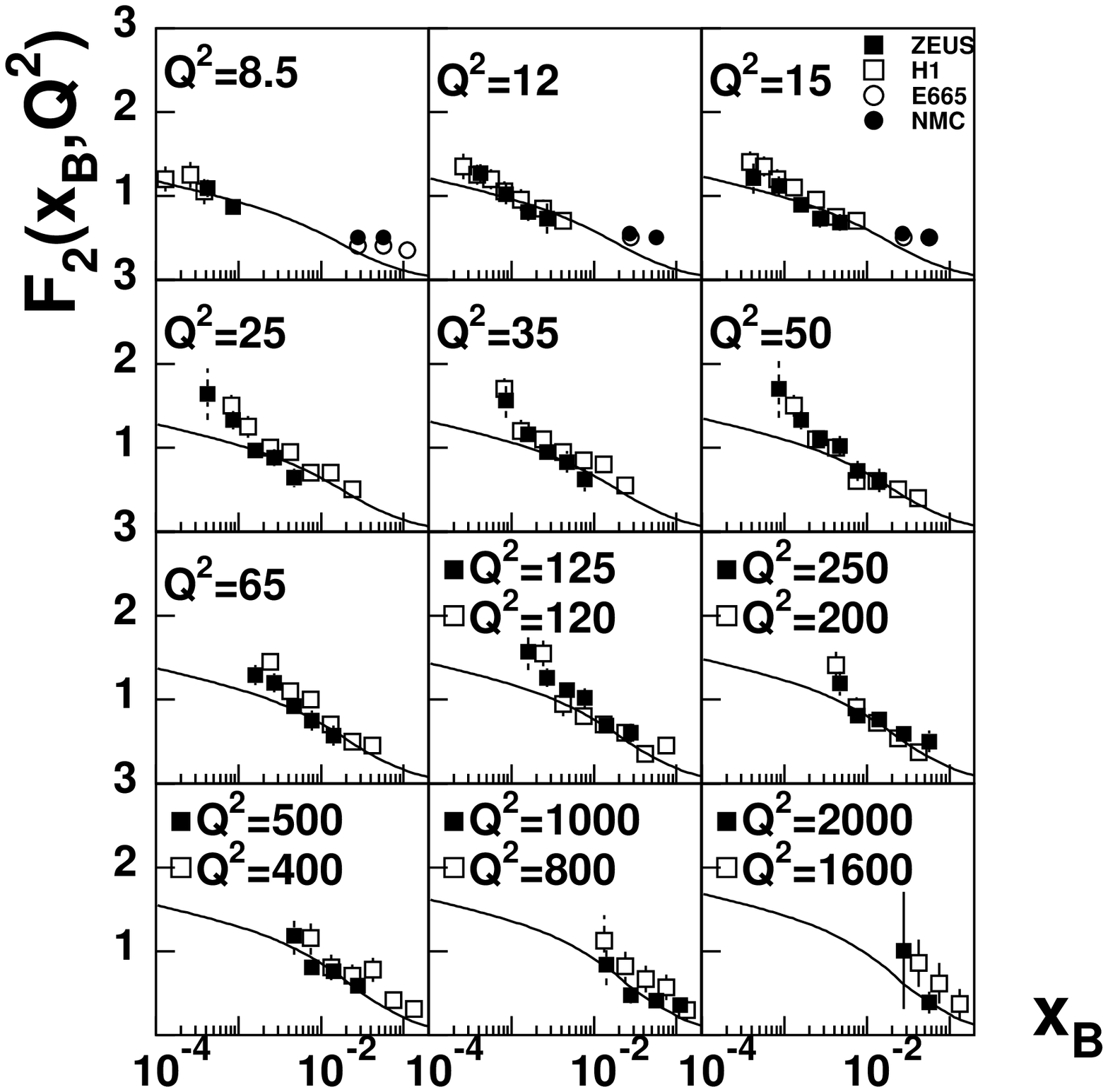,width=8cm}
\vspace*{-0.5cm}

\caption{ $F_2(x_B,Q^2)$, proton's structure function,  
(a) as function of $Q^2$
for different $x_B$-values, and (b)
as a function of $x_B$ for different $Q^2$-values.
The data points  are taken from Refs.[1,2,3,7].
The curves are the calculated results.
Here, as well as in Fig.3 and 4, the  lines
show the calculated result with the 
 $F_2^{ci}(x_{Bc})=$const.} 
\end{figure}

\begin{figure}
\psfig{file=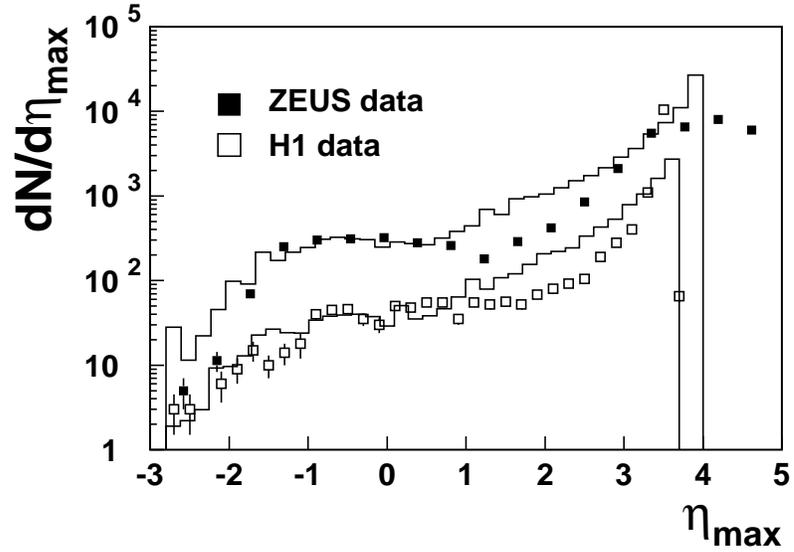,width=15cm,angle=270}
\caption{
The distribution of
$\eta_{max}$ compared with the data
(Refs. 1 and 2). Note that the
kinematic cuts used by H1 and ZEUS are different.}
\end{figure} 

\newpage

\begin{figure}
\psfig{file=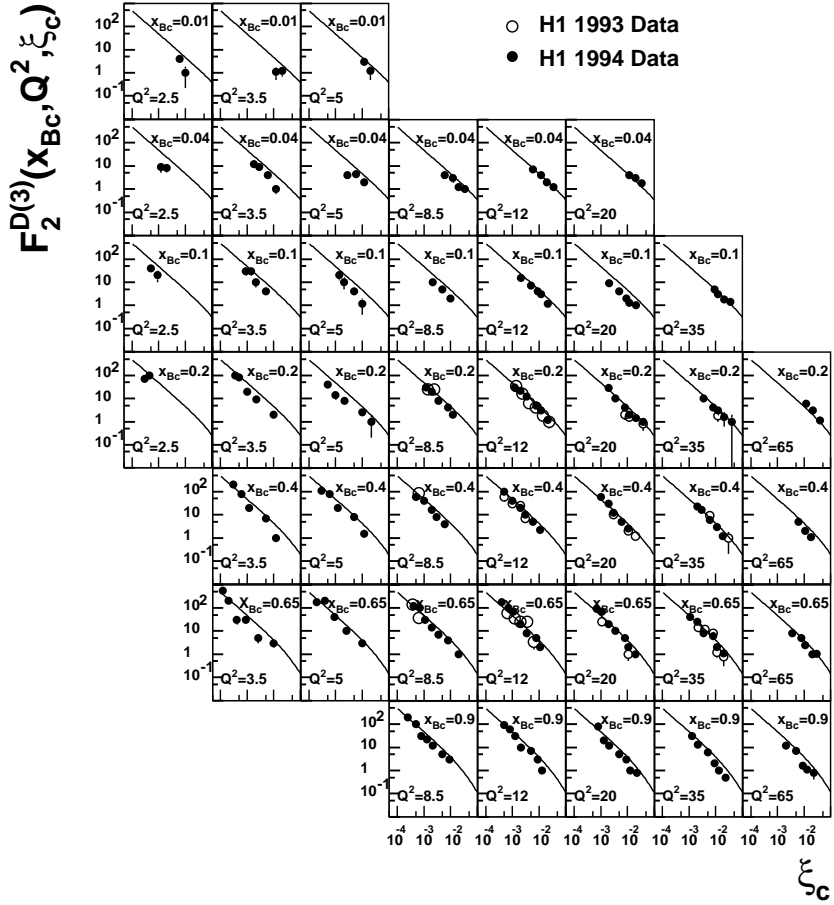,width=13cm}
\caption{
 $F_2^{D(3)}(x_{Bc},Q^2,\xi_c)$, the diffractive
structure function, 
as function of $\xi_c$ for different
values of $Q^2$ and 
$x_{Bc}$. The data are taken from Refs.1 and 3 
(Note that $\beta\equiv x_{Bc}$
and $x_P\equiv \xi_c$.) 
The lines are the calculated results.}
\end{figure} 
 
\end{document}